\documentclass{article}

\usepackage[utf8]{inputenc}
\usepackage[margin=30mm]{geometry}
\usepackage{parskip}
\usepackage{hyperref}
\usepackage{url}
\usepackage[pdftex]{graphicx}
\usepackage{textgreek}
\usepackage{booktabs}
\usepackage[square,numbers,compress]{natbib}
\usepackage{amsmath}
\usepackage{mathtools}
\usepackage{makecell}
\usepackage{arydshln}
\usepackage{pifont}
\usepackage{xcolor}

\usepackage[labelsep=space]{caption}
\newcommand{\tablepro}{\ding{51} }
\newcommand{\tablecon}{\ding{55} }
\newcommand{\revision}[1]{#1}

\title{Reversible molecular simulation for training classical and machine learning force fields}

\author{
  Joe G Greener \\
  Medical Research Council Laboratory of Molecular Biology \\
  Cambridge, UK \\
  \texttt{jgreener@mrc-lmb.cam.ac.uk} \\
}

\date{}

\begin{document}

\maketitle

\begin{abstract}
\normalsize
\noindent The next generation of force fields for molecular dynamics will be developed using a wealth of data.
Training systematically with experimental data remains a challenge, however, especially for machine learning potentials.
Differentiable molecular simulation calculates gradients of observables with respect to parameters through molecular dynamics trajectories.
Here we improve this approach by explicitly calculating gradients using a reverse-time simulation with effectively constant memory cost and a computation count similar to the forward simulation.
The method is applied to learn all-atom water and gas diffusion models with different functional forms, and to train a machine learning potential for diamond from scratch.
Comparison to ensemble reweighting indicates that reversible simulation can provide more accurate gradients and train to match time-dependent observables.
\end{abstract}

\section*{Introduction}

Molecular dynamics (MD) simulations have given us insight into how atoms move, from biomolecules to materials \cite{Hollingsworth2018}.
Key to the accuracy of a MD simulation is the accuracy of the force field used to describe how the atoms interact.
For classical molecular mechanics, force field development has largely been manual with parameters tuned to give the best possible match to quantum mechanical (QM) data (bottom-up) and condensed phase properties (top-down) \cite{Ding2023, Frohlking2020}.
There have been automated approaches, including ensemble reweighting methods \cite{Norgaard2008, DiPierro2013, Thaler2021, Fuchs2024, Navarro2023} like the popular ForceBalance \cite{Wang2014, Wang2013, Wang2013b}, and graph neural networks to avoid discrete atom typing \cite{Takaba2024}, but much work is still done manually \cite{Robustelli2018}.
The recently emerging and promising machine learning interatomic potentials (MLIPs) \cite{Smith2017, Batzner2022} are typically trained bottom-up on QM data alone \cite{Matin2024}, though this can give a distorted view of the utility of these models \cite{Fu2023}.
Whilst MLIPs can be validated on other data \cite{Kovacs2023}, using non-QM data during training has proved challenging.
This puts a heavy emphasis on generating large and diverse QM datasets and neglects other available data.

One approach to training force fields with experimental data is differentiable molecular simulation (DMS), in which automatic differentiation (AD) \cite{Baydin2018} is used to obtain the gradients of a loss value with respect to the parameters over a simulation.
This has had a number of recent applications \cite{Ingraham2019, Wang2020, Greener2021, Goodrich2021, Sipka2023, Wang2023, Greener2024, Gangan2024, King2024, Carrer2024, Wu2024, Han2025} with dedicated software available \cite{Doerr2021, Schoenholz2019, Wang2023b, Orlando2024, Greener2024}.
It is appealing due to the variety of possible loss functions and because the gradients are exact with respect to the forward simulation.
There are, however, three main problems with DMS.
Firstly, the memory required is linear in the number of simulation steps meaning that gradient checkpointing is required for longer simulations (reducing the memory scaling to logarithmic \cite{Chen2016}) and that larger neural networks may be incompatible.
Secondly, performance is considerably slower than standard simulation due to the overhead of reverse mode AD (RAD).
Finally, the gradients are prone to explosion due to the numerical integration.
\revision{DMS holds promise  despite this,} particularly for training on time-dependent observables where ensemble reweighting approaches are not generally applicable \cite{Bolhuis2024}.
Examples of these include diffusion coefficients, autocorrelation functions, relaxation rates, thermal conductivity and reaction rates, where available data is challenging to use during training.

Here we take inspiration from reversible differential equation solvers \cite{Kidger2021, Sapienza2024, McCallum2024} and reversible neural networks \cite{Gomez2017, Chang2017, Maclaurin2015} and ask if DMS can be done without storing intermediate states, i.e.\ by explicitly deriving gradients rather than using conventional AD.
This is motivated by three features of molecular simulations: they consist of the same step repeated many times, the algorithm does not contain branching, and they are reversible in certain situations.
We find that identical gradients to DMS with RAD can be obtained with effectively constant memory cost and a computation count comparable to standard simulation, and explore gradient truncation as a way to avoid gradient explosion.
This reversible simulation approach is demonstrated with three examples: learning molecular mechanics water models with different functional forms, training to match gas diffusion data, and learning a MLIP for diamond from scratch.

\section*{Results}

\subsubsection*{Reversible molecular simulation}

\begin{figure}
  \centering
  \includegraphics[width=1.0\textwidth]{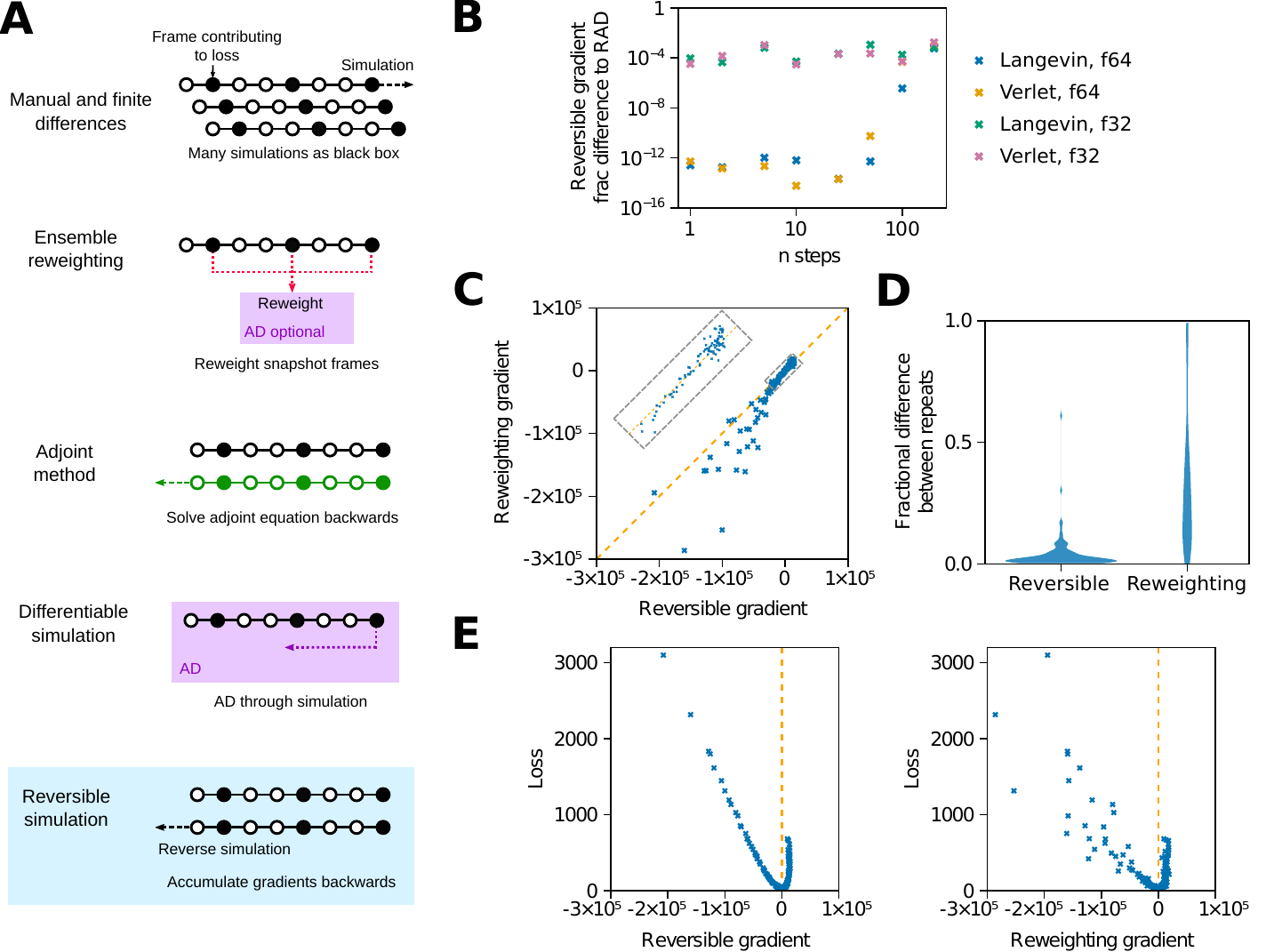}
  \caption{Reversible molecular simulation. (A) Different approaches to calculate gradients through molecular simulations. See also Table~\ref{tab:approaches}. (B) The difference between gradients obtained over simulations of TIP3P water from reversible simulation and DMS with RAD, which is mathematically equivalent but shows discrepancies due to a different order of floating point operations. \revision{Fractional differences} are shown for the Langevin middle integrator with $\gamma = 1$ ps\textsuperscript{-1} and double precision floats (f64) used throughout the work, plus the leapfrog Verlet integrator and single precision floats (f32). (C) Correlation of the gradients from reversible simulation and ensemble reweighting for simulations with the same parameters. For C-E the Lennard-Jones $\sigma$ gradient from 50 ps simulations of the water model is shown, with the simulation setup matching the water model training. To show a variety of loss and gradient values, 125 sets of parameters were run corresponding to $\sigma$ in (0.301, 0.308, 0.315, 0.322, 0.329) nm, $\varepsilon$ in (0.436, 0.536, 0.636, 0.736, 0.836) kJ/mol and O partial charge in (-0.934, -0.884, -0.834, -0.784, -0.734). (D) Testing how gradients change with repeats. The distribution of fractional differences in gradients from pairs of runs with the same parameters but different random seeds \revision{for the thermostat} is shown. (E) The loss values plotted against the gradients for each parameter set.}
  \label{fig:gradients}
\end{figure}

A molecular simulation is run using a force field with parameters $\sigma_{j}$.
We wish to improve $\sigma_{j}$ to better match experimental data.
Whilst it is possible to do this using gradient-free approaches, this scales poorly with parameter number and both molecular mechanics force fields and MLIPs can have thousands \revision{of parameters or more}.
Consequently, we wish to calculate $\frac{\textrm{d} l}{\textrm{d} \sigma_{j}}$ where the loss function $l$ represents the match of the simulation to experiment.
Existing gradient-based approaches to parameterise force fields are summarised in Table~\ref{tab:approaches} and Figure~\ref{fig:gradients}A.

\begingroup
\renewcommand{\arraystretch}{2.5} 
\begin{table}
  \centering
  \begin{small}
    \begin{tabular}{ l l l l }
      \hline
      \textbf{Class} & \textbf{Method} & \textbf{Pros and cons} \\
      \hline
      Not gradient-based & Manual adjustment \cite{Robustelli2018} & \makecell[l]{\tablepro Use human expertise \\ \tablepro Use fast software \\ \tablecon Poor scaling in parameter number \\ \tablecon Takes human time} \\
        & Sampling approaches \cite{Kummerer2023} & \makecell[l]{\tablepro Automated \\ \tablepro Use fast software \\ \tablecon Poor scaling in parameter number} \\
      \hdashline
      Numerical & Finite differences & \makecell[l]{\tablepro Accurate gradients for smooth functions \\ \tablepro Use fast software \\ \tablecon Poor scaling in parameter number \\ \tablecon Needs tuning or can be inaccurate} \\
      \hdashline
      Ensemble-based & \makecell[l]{Ensemble reweighting, \\ e.g.\ ForceBalance \cite{Wang2014}} & \makecell[l]{\tablepro Applicable to a variety of properties \\ \tablepro Can be enhanced with AD \\ \tablecon Not applicable to time-dependent properties} \\
       & \makecell[l]{Differentiable trajectory \\ reweighting (DiffTre) \cite{Thaler2021}} & \makecell[l]{\tablepro More efficient version of ensemble reweighting \\ \tablecon Not applicable to time-dependent properties} \\
       \hdashline
      Trajectory-based & Adjoint method \cite{Chen2018} & \makecell[l]{\tablepro Memory efficient \\ \tablecon Solves separate adjoint equation \\ \tablecon Can be unstable} \\
       & Differentiable simulation \cite{Greener2024} & \makecell[l]{\tablepro Accurate gradients \\ \tablecon Poor memory scaling with trajectory length} \\
       & \makecell[l]{Reversible simulation \\ (this work)} & \makecell[l]{\tablepro Memory-efficient \\ \tablecon Needs custom implementation} \\
      \hline
    \end{tabular}
  \end{small}
  \caption{Approaches to parameterise force fields, with a focus on methods that calculate the gradient of the match to experiment with respect to the force field parameters $\frac{\textrm{d} l}{\textrm{d} \sigma_{j}}$. See also Figure~\ref{fig:gradients}A.}
  \label{tab:approaches}
\end{table}
\endgroup

Here we show (see the Methods) that:
\begin{align*}
\frac{\mathrm{d} \langle l \rangle}{\mathrm{d} \sigma_{j}} &= \left\langle \frac{\partial l}{\partial \sigma_{j}} \right\rangle + \left\langle \sum_{i = 1}^{n_s - 1} \frac{\mathrm{d} l}{\mathrm{d} \mathbf{f}_{i}}^\top \frac{\partial F(\mathbf{x}_{i}, \sigma_{j})}{\partial \sigma_{j}} \right\rangle \tag{1}
\end{align*}
where $\mathbf{x}_{i}$ are the coordinates at step $i$, $\mathbf{f}_{i}$ are the forces on each atom at step $i$, $F$ is the force function, $n_s$ is the snapshot step, and the angle brackets represent the average over \revision{loss} snapshots of the simulation.
$\frac{\partial F(\mathbf{x}_{i}, \sigma_{j})}{\partial \sigma_{j}}$ can be calculated at each time step.
By calculating a series of intermediate values, $\frac{\mathrm{d} l}{\mathrm{d} \mathbf{f}_{i}}$ can be accumulated by stepping back in time.
This equates to the same operations as DMS with RAD but coded explicitly, and requires running the simulation back in time, hence the name reversible simulation.

Arbitrary trajectories back in time will diverge in the NVT (canonical) ensemble, hence an initial simulation forwards in time must be run for the length of the simulation to ensure we obtain a valid trajectory.
Given the tendency of the reverse-time integrator to gradually diverge over time from the corresponding forward simulation due to not being bitwise reversible \cite{Levesque1993, Kidger2021b, Maclaurin2015}, snapshots also need to be stored every 1 ps to reset the coordinates and velocities.
Apart from this storage, which is cheap, the method is constant in memory for any simulation length.
Conceptually it is similar to the adjoint method \cite{Chen2018, Han2025}, with a comparison in the Methods, though the adjoint method solves a different equation back in time \cite{Gholami2019, Onken2020, Kim2021, Norcliffe2020}.

\subsubsection*{Learning water models}

\begin{figure}
  \centering
  \includegraphics[width=0.85\textwidth]{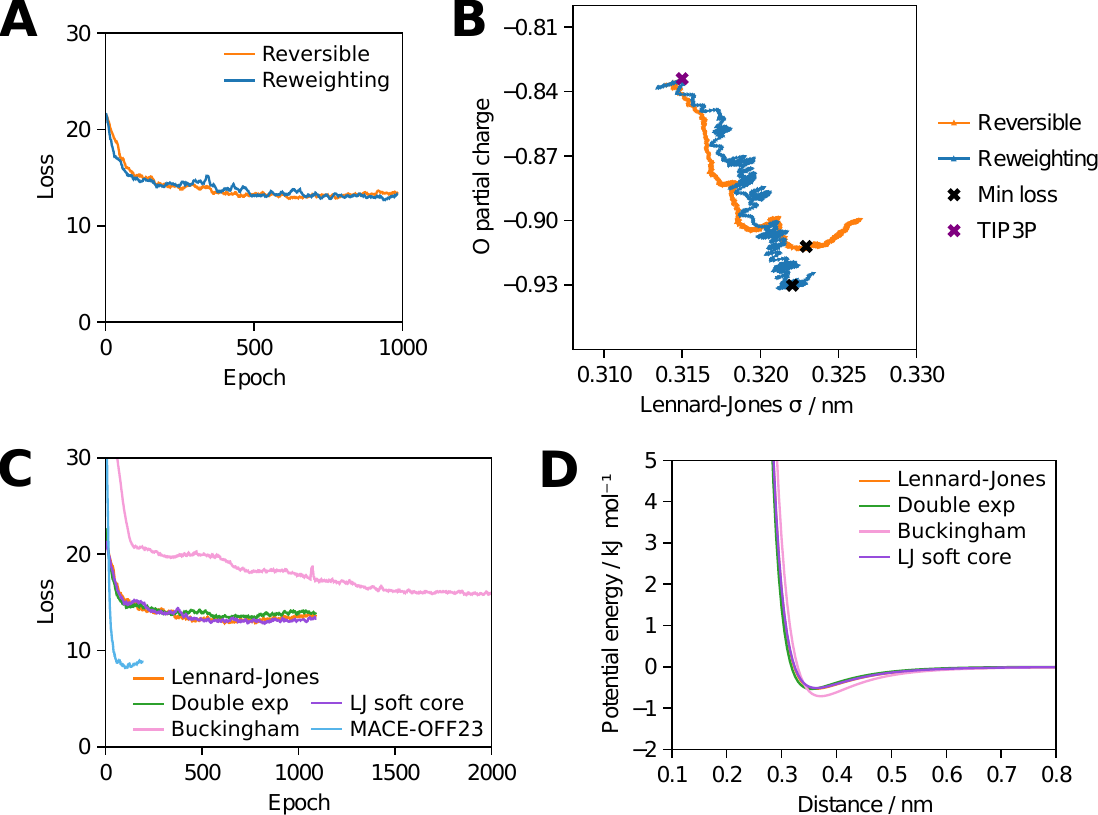}
  \caption{Training a 3-point water model. A and C show values smoothed by taking the mean of values up to 5 places either side. (A) Progress over training for reversible simulation and ensemble reweighting using a loss based on enthalpy of vapourisation and RDF. Each epoch, gradients from a 50 ps simulation were used to update the parameters. (B) The change in the Lennard-Jones $\sigma$ and O partial charge parameters during training. (C) Training different functional forms with reversible simulation. The Lennard-Jones case is the same as in A. The Buckingham case was trained for longer due to lack of convergence. \revision{The MACE-OFF23 model was fine-tuned from the pre-trained model.} (D) The shape of the non-charge non-bonded potential energy function for the trained parameters of each functional form.}
  \label{fig:water_train}
\end{figure}

\begin{figure}
  \centering
  \includegraphics[width=1.0\textwidth]{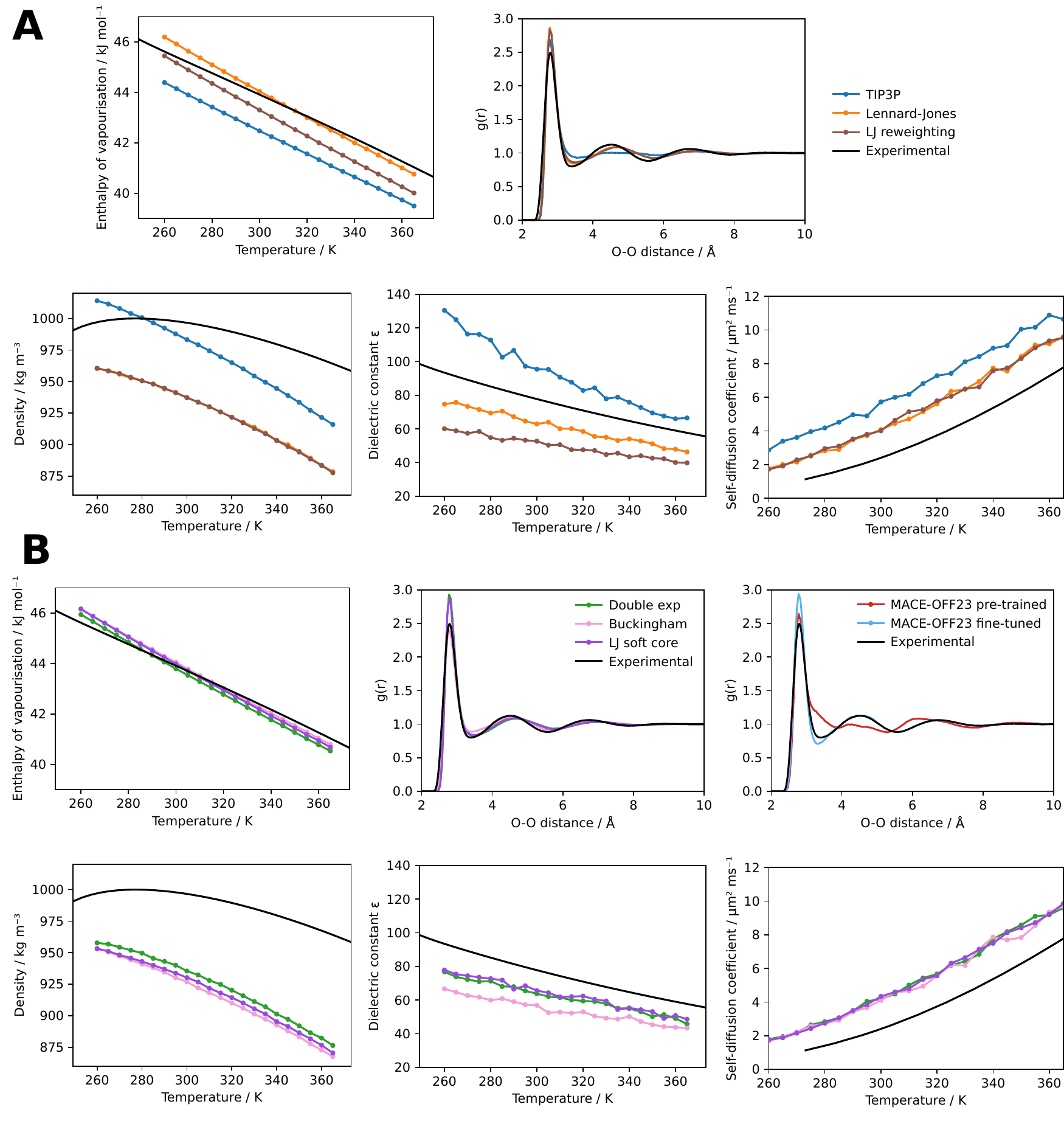}
  \caption{\revision{Validating water models on condensed phase properties. Experimental values are taken from Wang et al.\ 2014 \cite{Wang2014}, Soper 2013 \cite{Soper2013} for the RDF and Easteal et al.\ 1989 \cite{Easteal1989} for the self-diffusion coefficient. The RDF is calculated from a simulation at 300 K. (A) Properties for the Lennard-Jones parameters trained with reversible simulation, plus the Lennard-Jones parameters trained with ensemble reweighting and the TIP3P starting parameters. (B) Properties for other functional forms trained with reversible simulation. The RDF with MACE-OFF23 before and after fine-tuning with reversible simulation is also shown.}}
  \label{fig:water_val}
\end{figure}

To test reversible simulation and compare to ensemble reweighting, we train a 3-point molecular mechanics water model to match experimental data.
Parameterising water models is a common challenge where the fit to various properties has to be balanced.
In this case enthalpy of vapourisation and radial distribution function (RDF) \cite{Wade2018} data were used as a proof of principle, though other studies have used more properties \cite{Wang2014}.
Starting from the popular TIP3P water model \cite{Jorgensen1983} we train the Lennard-Jones $\sigma$ and $\varepsilon$ parameters, the partial charge on oxygen (and hence on hydrogen, since the overall molecular charge is zero), and the equilibrium values and force constants for the harmonic bonds and angles.
As can be seen in Figure~\ref{fig:gradients}B, the gradients are numerically identical to DMS with RAD for small step numbers as expected.
The gradients from reversible simulation correlate surprisingly well with those from ensemble reweighting, which are compared to in Figure~\ref{fig:gradients}C.
It is encouraging that these two distinct approaches give similar gradients.
The gradients vary much less for reversible simulation over repeats with different random seeds used for the thermostat (Figure~\ref{fig:gradients}D).
This is possibly due to the increased number of steps contributing to the gradient as discussed in the Supplementary Methods.
Plotting the loss values against the gradients shows that the loss is minimised when the gradient is zero, indicating that the gradients are accurate and that optimising with the gradients will minimise the loss (Figure~\ref{fig:gradients}E).
The correlation of loss to gradient magnitude is better for reversible simulation, suggesting that it may provide a smoother optimisation surface.

As shown in Figure~\ref{fig:water_train}A both reversible simulation and ensemble reweighting provide gradients that improve the match to experiment for the chosen properties over training with simulations of 50 ps using a box of 895 water molecules.
They follow similar optimisation pathways through parameter space, shown in Figure~\ref{fig:water_train}B for two parameters, with reversible simulation taking steps in a more consistent direction than ensemble reweighting as suggested by Figure~\ref{fig:gradients}E.
Longer validation simulations with the learned potentials show an improved match to the enthalpy of vapourisation across multiple temperatures and to the RDF (Figure~\ref{fig:water_val}), though ensemble reweighting does not match the enthalpy of vapourisation as well.
Other properties are also shown.
The match to density is made worse as it was not used during training, though the match to the self-diffusion coefficient is improved.

Rather than fit to all available properties here, we aim to demonstrate that reversible simulation is able to match chosen experimental properties for all-atom models.
\revision{In principle any number of loss functions can be combined with reversible simulation, with a weighting applied to each.
This opens up opportunities to fit to properties of interest without making the fit to existing properties worse.
The ability to fit to multiple properties depends on the expressive power of the force field, and molecular mechanics 3-point water models for example are known to struggle to fit all available structural and dynamic properties \cite{Izadi2014}.}

\revision{\subsubsection*{Alternative functional forms}}

Since reversible simulation is independent of the functional form used to calculate the forces, we also demonstrate that it can optimise parameters for other functional forms of the non-charge non-bonded potential.
The double exponential, Buckingham and Lennard-Jones soft core potentials have all been proposed as improvements over the Lennard-Jones potential, in which the repulsion term is not physically motivated.
By starting from sensible parameters and training on the same properties as before, parameters can be learned that better fit the experimental data.
As can be seen in Figure~\ref{fig:water_train}C-D and Figure~\ref{fig:water_val}B these flexible functional forms give potentials of a similar shape with the learned parameters and are able to match the enthalpy of vapourisation and RDF well.
This indicates that reversible simulation could be useful in developing the next generation of force fields that go beyond Lennard-Jones.

\revision{The MACE-OFF23 MLIP has been developed for organic molecules but shows some discrepancies to experiment for water properties \cite{Kovacs2023}.
We fine-tuned the pre-trained small MACE-OFF23 model in the same manner as the molecular mechanics force fields above.
As shown in Figure~\ref{fig:water_train}C and Figure~\ref{fig:water_val}B this gives a better match to experiment, with the higher expressivity of the MLIP giving a lower loss than the molecular mechanics cases.
Validation simulations at 300 K show an improved match of the RDF to experiment and the enthalpy of vapourisation improves from the pre-trained value of 49.4 kJ/mol to 44.9 kJ/mol, closer to the experimental value of 43.9 kJ/mol.
The density also improves from the pre-trained value of 1116 kg/m\textsuperscript{3} to 992 kg/m\textsuperscript{3}, closer to the experimental value of 997 kg/m\textsuperscript{3}.
This indicates that reversible simulation is able to fine-tune larger models, in this case one with 694,320 parameters, to match experimental data.
Ensemble reweighting can also be used to fine-tune the model, giving a similar RDF but a worse match for the enthalpy of vapourisation (47.7 kJ/mol) compared to the model fine-tuned with reversible simulation.}

As discussed in the Methods, the run time of reversible simulation is similar to that of the forward simulation.
For water training \revision{(2685 atoms)} the run time was 2.7 ms per simulation step on CPU for Lennard-Jones, compared to 2.3 ms for a single forward step.
In comparison the run time of OpenMM on the same system was 1.2 ms per step on CPU for a standard simulation, so reversible simulation can approach the simulation speed of mature software.
Optimisation for \revision{the} GPU is left to further work.
The alternative functional forms add less than 10\% to the run time of Lennard-Jones.
\revision{MACE-OFF23 takes 180 ms per simulation step on one A100 GPU, compared to 120 ms for a single forward step.}

\subsubsection*{Gas diffusion in water}

\begin{figure}
  \centering
  \includegraphics[width=0.95\textwidth]{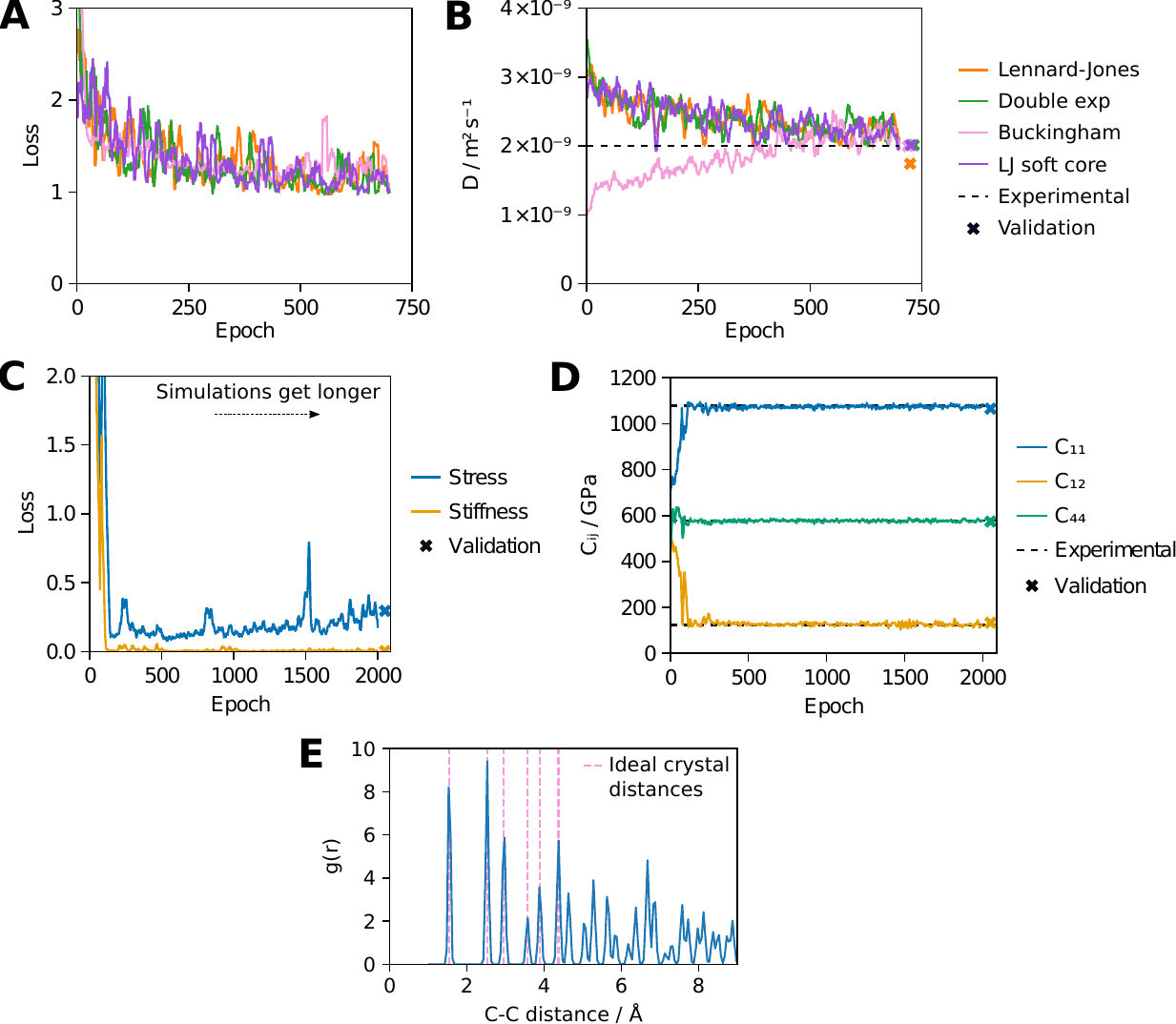}
  \caption{Gas diffusion and diamond models. \revision{A-D} show values smoothed by taking the mean of values up to 5 places either side. (A) Gas diffusion model. The loss, \revision{consisting of the match of the diffusion coefficient $D$ to experiment plus the losses from the water model,} over \revision{700} epochs of training is shown for each functional form. (B) $D$ calculated from the simulation at each epoch, compared to the experimental $D$ value of $2.0 \times 10^{-9}$ m\textsuperscript{2} s\textsuperscript{-1} at the simulation temperature \cite{Han1996}. The $D$ value \revision{with the final trained model} averaged over 5 validation simulations of 100 ps is also shown. (C) Diamond model. The stress ($\boldsymbol\sigma^V$) and stiffness ($\mathbf{C}$) contributions to the loss are shown for 2000 epochs of training, with the simulation length increasing linearly each step and ending at 1 ps. The loss for validation simulations of 100 ps is also shown. (D) The 3 distinct stiffness moduli contributing to the stiffness tensor $\mathbf{C}$ \cite{Thaler2021} shown over training, along with the values for validation simulations and the experimental values. \revision{(E) The RDF of carbon atoms in diamond for a simulation with the trained model. The distances in the ideal diamond cubic crystal are also shown.}}
  \label{fig:diff_diamond}
\end{figure}

Given that ensemble reweighting gives similar gradients to reversible simulation (Figure~\ref{fig:gradients}C) and is often easier to set up, it will be the preferred choice for many properties of interest.
However, reversible simulation is distinguished by its ability to target time-dependent properties.
Here we show how this can be useful by learning parameters that match the experimental diffusion coefficient $D$ of the oxygen diatomic molecule in water.
For Lennard-Jones we use TIP3P starting parameters for the water and oxygen parameters from Wang et al.\ 2021 \cite{Wang2021}.
\revision{The loss was the deviation from the experimental $D$ value, plus the enthalpy of vapourisation and RDF losses from the water model to prevent degradation in the bulk water properties.}
By training on simulations of 50 ps with 10 oxygen molecules randomly placed in 885 water molecules and calculating $D$ using the slope of the mean squared displacement (MSD) against time, reversible simulation can learn parameters that reproduce the experimental value of $2.0 \times 10^{-9}$ m\textsuperscript{2} s\textsuperscript{-1} for $D$ from a starting value of $3.6 \times 10^{-9}$ m\textsuperscript{2} s\textsuperscript{-1} (Figure~\ref{fig:diff_diamond}A-B).

Similar to the water models discussed previously, we learn parameters for alternative functional forms.
These are also able to reproduce the experimental value of $D$, indicating that reversible simulation can train to match time-dependent properties for a variety of functional forms.
Longer simulations with the learned parameters reproduce improved $D$ values, as shown in Figure~\ref{fig:diff_diamond}B.
\revision{The learned parameters also give condensed phase properties for bulk water similar to Figure~\ref{fig:water_val}, as shown in Figure~\ref{fig:diff_val}, demonstrating that dynamic properties can be targeted without degrading the fit to other properties.}

\subsubsection*{Neural network model for diamond}

The \revision{molecular mechanics} water molecules described above have fewer than 10 parameters each.
In order to demonstrate that reversible simulation can train neural networks with many more parameters from scratch, we train the MLIP model for diamond used in Thaler and Zavadlav 2021 \cite{Thaler2021} on \revision{a} GPU using the experimental elastic stiffness tensor.
The model consists of a Stillinger-Weber prior with starting parameters for silicon \cite{Stillinger1985} and the DimeNet++ neural network \cite{Gasteiger2020}.
The virial stress tensor and stiffness tensor calculated via the stress fluctuation method were used to define the loss function, with only three distinct stiffness moduli in the stiffness tensor due to symmetries in the diamond cubic crystal.

All parameters of the model were trained over increasing numbers of simulation steps of 1000 carbon atoms, with 1 ps of simulation used by the end of training.
This was sufficient to train the model, as shown in Figure~\ref{fig:diff_diamond}C-D.
The learned model maintains a low loss over longer 100 ps validation simulations, indicating stability, with stress and stiffness values showing good agreement with the target values.
\revision{Figure~\ref{fig:diff_diamond}E shows that the RDF from simulations with the trained model aligns with the expected structural properties of the diamond cubic crystal.}
The DimeNet++ model used has 121,542 parameters, demonstrating that reversible simulation can effectively train models with large numbers of parameters \revision{from scratch}.
As long as one force evaluation can fit in memory, reversible simulation should be applicable to even larger models whereas DMS with RAD would struggle even with gradient checkpointing.

\section*{Discussion}

The number of computations required to calculate gradients with reversible simulation is similar to that of a standard simulation due to gradient truncation, as described in the Methods.
In addition, reversible simulation uses effectively constant memory for any number of simulation steps, is applicable to many loss functions and gives accurate gradients that numerically match those from the forward simulation, unlike the adjoint method.
The high degree of control over the gradients, not available in general with AD, means that gradient truncation can be easily implemented.
These improvements over DMS with RAD should make it applicable to larger systems and systems where the potential has a significant memory cost such as MLIPs.
The ability to train three different systems using the Adam optimiser with gradients from reversible simulation shows its wide applicability.

One drawback of the method is that it requires implementing the algorithm whereas ensemble reweighting can largely make use of existing software.
However, implementing the algorithm is not particularly difficult and can mostly be achieved using fast components of existing software.
Another drawback is that the loss and force functions need to be differentiable with respect to the atomic coordinates, which can be challenging for losses such as density.
Second order AD may be required to calculate the force gradients for MLIPs, but this is supported in many frameworks.
Some loss functions involving combinations of averages are also hard to implement with reversible simulation.

Recent work has used neural networks for continuous atom typing \cite{Takaba2024}.
These methods could be trained end-to-end with reversible simulation to target condensed phase properties.
It should also be possible to train on binding free energy data directly \cite{Wang2023b, Setiadi2024} with reversible simulation by differentiating through the appropriate estimator.
One surprise from this work is the similarity between gradients arising from reversible simulation and ensemble reweighting.
This is encouraging given that they are computed in different ways.
For many applications, ensemble-based approaches are sufficient.
However, reversible simulation allows time-dependent properties to be targeted and here gives gradients with less variance.
It could be used in combination with ensemble reweighting to target multiple properties, alongside force matching to QM data \cite{Rocken2024}.
A variety of approaches will be important for training the next generation of molecular mechanics force fields, MLIPs, and everything in-between \cite{Wang2024}.

\section*{Methods}

\subsubsection*{Reversible molecular simulation}

Consider the widely used Langevin integrator for running molecular simulations in the NVT (canonical) ensemble:
$$
\textbf{m} \textbf{a}_{i} = F(\mathbf{x}_{i}, \sigma_{j}) - \gamma \textbf{m} \textbf{v}_{i} + \sqrt{2 \mathbf{m} \gamma k_{B} T} \textbf{R}_{i}
$$
where for $N$ atoms at step $i$, $\mathbf{x}_{i}$ are the atomic coordinates, $\mathbf{v}_{i}$ are the velocities, $\mathbf{a}_{i}$ are the accelerations, $\mathbf{m}$ are the masses, $F$ is the force function arising from the interaction potential, $\sigma_{j}$ are the force field parameters, $\gamma$ is the collision frequency, $k_{B}$ is the Boltzmann constant, $T$ is the temperature and $\textbf{R}_{i}$ is a stationary Gaussian process with zero-mean.
One popular implementation is the Langevin middle integrator from OpenMM \cite{Zhang2019, Eastman2017}, which has been used successfully for DMS \cite{Greener2024}.
The integration step at step $i$ for this integrator is:
\begin{align*}
\textbf{f}_{i} &= F(\mathbf{x}_{i}, \sigma_{j}) \\
\textbf{a}_{i}^{'} &= \textbf{f}_{i} / \textbf{m} \\
\mathbf{v}_{i+1/2}^{'} &= \mathbf{v}_{i-1/2} + \Delta t \textbf{a}_{i}^{'} \\
\mathbf{x}_{i+1/2}^{'} &= \mathbf{x}_{i} + \frac{\Delta t}{2} \mathbf{v}_{i+1/2}^{'} \\
\mathbf{v}_{i+1/2} &= e^{-\gamma \Delta t} \mathbf{v}_{i+1/2}^{'}  + \sqrt{1 - e^{-2 \gamma \Delta t}} \textbf{n}_{i} \\
\mathbf{x}_{i+1} &= \mathbf{x}_{i+1/2}^{'} + \frac{\Delta t}{2} \mathbf{v}_{i+1/2} \tag{2}
\end{align*}
where $\textbf{f}_{i}$ are the forces arising from the interaction potential, $\Delta t$ is the time step, $\mathbf{n}_{i}$ are random velocities generated from the Boltzmann distribution at temperature $T$ each step and $'$ denotes intermediate computation values.
The velocities are offset by half a time step from the coordinates.
If the match to experiment after a simulation of $n$ steps is represented by a loss function $l(\mathbf{x}_{n}, \mathbf{v}_{n-1/2}, \sigma_{j})$ then according to the multi-variable chain rule:
$$
\frac{\mathrm{d} l}{\mathrm{d} \sigma_{j}} = \frac{\partial l}{\partial \sigma_{j}} + \sum_{i = 1}^{n - 1} \frac{\mathrm{d} l}{\mathrm{d} \mathbf{f}_{i}}^\top \frac{\partial F(\mathbf{x}_{i}, \sigma_{j})}{\partial \sigma_{j}}
$$
since $\sigma_{j}$ only appears in $l$ and $F$ during the integration step (Equation~2).
In the case that multiple snapshots contribute to the loss, then:
$$
\frac{\mathrm{d} \langle l \rangle}{\mathrm{d} \sigma_{j}} = \left\langle \frac{\partial l}{\partial \sigma_{j}} \right\rangle + \left\langle \sum_{i = 1}^{n_s - 1} \frac{\mathrm{d} l}{\mathrm{d} \mathbf{f}_{i}}^\top \frac{\partial F(\mathbf{x}_{i}, \sigma_{j})}{\partial \sigma_{j}} \right\rangle
$$
where $n_s$ is the step number of the snapshot and the angle brackets represent the average over the snapshots.
$\frac{\partial l}{\partial \sigma_{j}}$ can be calculated at the point of calculating $l$.
$\frac{\partial F(\mathbf{x}_{i}, \sigma_{j})}{\partial \sigma_{j}}$ can be calculated each step, shown in the Supplementary Methods for the example of the Lennard-Jones potential, meaning that the challenge is to calculate the $\frac{\mathrm{d} l}{\mathrm{d} \mathbf{f}_{i}}$ terms.
This can be rewritten:
\begin{align*}
\frac{\mathrm{d} l}{\mathrm{d} \mathbf{f}_{i}} &= \frac{\mathrm{d} l}{\mathrm{d} \mathbf{x}_{n}}^\top \frac{\mathrm{d} \mathbf{x}_{n}}{\mathrm{d} \mathbf{f}_{i}} + \frac{\mathrm{d} l}{\mathrm{d} \mathbf{v}_{n-1/2}}^\top \frac{\mathrm{d} \mathbf{v}_{n-1/2}}{\mathrm{d} \mathbf{f}_{i}} \tag{3}
\end{align*}
The terms can be derived using Symbolics.jl \cite{Gowda2022} from an unrolled simulation (see the Supplementary Methods).
The first two terms are:
\begin{align*}
\frac{\mathrm{d} l}{\mathrm{d} \mathbf{f}_{n-1}} &= \frac{\Delta t^{2}}{2 \mathbf{m}} \bigg ( 1 + e^{-\gamma \Delta t} \bigg ) \frac{\mathrm{d} l}{\mathrm{d} \mathbf{x}_{n}} + \frac{\Delta t}{\mathbf{m}} e^{-\gamma \Delta t} \frac{\mathrm{d} l}{\mathrm{d} \mathbf{v}_{n-1/2}} \\
\frac{\mathrm{d} l}{\mathrm{d} \mathbf{f}_{n-2}} &= \begin{multlined}[t] \frac{\Delta t^{2}}{2 \mathbf{m}} \bigg ( 1 + 2 e^{-\gamma \Delta t} + e^{-2 \gamma \Delta t} \bigg ) \bigg ( \frac{\mathrm{d} l}{\mathrm{d} \mathbf{x}_{n}} + \frac{\Delta t^{2}}{2 \mathbf{m}} \frac{\mathrm{d} l}{\mathrm{d} \mathbf{x}_{n}}^\top \frac{\mathrm{d} F(\mathbf{x}_{n-1}, \sigma_{j})}{\mathrm{d} \mathbf{x}_{n-1}} \bigg ) \\ + \frac{\Delta t}{\mathbf{m}} \bigg ( e^{-2 \gamma \Delta t} \frac{\mathrm{d} l}{\mathrm{d} \mathbf{v}_{n-1/2}} + \frac{\Delta t^{2}}{2 \mathbf{m}} \bigg ( e^{-\gamma \Delta t} + e^{-2 \gamma \Delta t} \bigg ) \frac{\mathrm{d} l}{\mathrm{d} \mathbf{v}_{n-1/2}}^\top \frac{\mathrm{d} F(\mathbf{x}_{n-1}, \sigma_{j})}{\mathrm{d} \mathbf{x}_{n-1}} \bigg ) \end{multlined} \tag{4}
\end{align*}
Noting that $\frac{\mathrm{d} l}{\mathrm{d} \mathbf{f}_{i}}$ accumulates terms for each step backwards in time, this suggests an efficient approach to calculating $\frac{\mathrm{d} l}{\mathrm{d} \sigma_{j}}$ by running a reverse-time simulation.
This is mathematically equivalent to RAD.
The concept is similar to using a reversible differential equation solver \cite{Kidger2021, Sapienza2024, McCallum2024} and reversible neural networks \cite{Gomez2017, Chang2017, Maclaurin2015}, with a discussion in Section 5.3.2 of Kidger 2021 \cite{Kidger2021}.
For the Langevin middle integrator, the time step is reversible provided that the random velocities from the previous step, $\mathbf{n}_{i-1}$, are known:
\begin{align*}
\mathbf{x}_{i-1/2}^{'} &= \mathbf{x}_{i} - \frac{\Delta t}{2} \mathbf{v}_{i-1/2} \\
\mathbf{v}_{i-1/2}^{'} &= e^{-\gamma \Delta t} \bigg ( \mathbf{v}_{i-1/2} - \sqrt{1 - e^{-2 \gamma \Delta t}} \textbf{n}_{i-1} \bigg ) \\
\mathbf{x}_{i-1} &= \mathbf{x}_{i-1/2}^{'} - \frac{\Delta t}{2} \mathbf{v}_{i-1/2}^{'} \\
\textbf{f}_{i-1} &= F(\mathbf{x}_{i-1}, \sigma_{j}) \\
\textbf{a}_{i-1}^{'} &= \textbf{f}_{i-1} / \textbf{m} \\
\mathbf{v}_{i-3/2} &= \mathbf{v}_{i-1/2}^{'} - \Delta t \textbf{a}_{i-1}^{'}
\end{align*}
Note that this integrator is not bitwise reversible \cite{Levesque1993, Kidger2021b, Maclaurin2015} since the order of floating point operations is different to the forward step.
Consequently, coordinates and velocities are stored every 1 ps and reset during the reverse simulation to prevent drift.
This incurs a small memory cost proportional to the number of simulation steps.
A series of accumulation vectors is required to update $\frac{\mathrm{d} l}{\mathrm{d} \mathbf{f}_{i}}$.
The starting values at step $n$ are:
\begin{align*}
\mathbf{A}_{n} &= \bigg ( 1 + e^{-\gamma \Delta t} \bigg ) \frac{\mathrm{d} l}{\mathrm{d} \mathbf{x}_{n}} + \frac{2}{\Delta t} e^{-\gamma \Delta t} \frac{\mathrm{d} l}{\mathrm{d} \mathbf{v}_{n-1/2}} \\
\mathbf{B}_{n} &= \bigg ( 1 + e^{-\gamma \Delta t} \bigg ) \frac{\mathrm{d} l}{\mathrm{d} \mathbf{x}_{n}} + \frac{2}{\Delta t} \bigg ( e^{-\gamma \Delta t} - 1 \bigg ) \frac{\mathrm{d} l}{\mathrm{d} \mathbf{v}_{n-1/2}} \\
\mathbf{C}_{n} &= \frac{1}{\Delta t} \bigg ( 1 + e^{-\gamma \Delta t} \bigg ) \frac{\mathrm{d} l}{\mathrm{d} \mathbf{v}_{n-1/2}} \\
\mathbf{D}_{n} &= \frac{1}{2} \bigg ( e^{\gamma \Delta t} + 2 + e^{-\gamma \Delta t} \bigg ) \frac{\mathrm{d} l}{\mathrm{d} \mathbf{x}_{n}} + \frac{1}{\Delta t} \bigg ( e^{-\gamma \Delta t} - e^{\gamma \Delta t} \bigg ) \frac{\mathrm{d} l}{\mathrm{d} \mathbf{v}_{n-1/2}}
\end{align*}
At each time step, the accumulation vectors, $\frac{\mathrm{d} l}{\mathrm{d} \mathbf{f}_{i}}$ and the growing $\frac{\mathrm{d} l}{\mathrm{d} \sigma_{j}}$ are updated:
\begin{align*}
\mathbf{D}_{i-1} &= e^{-\gamma \Delta t} \mathbf{D}_{i} + \bigg ( 1 + e^{-\gamma \Delta t} \bigg ) \frac{\Delta t^{2}}{2} \frac{\mathrm{d} l}{\mathrm{d} \mathbf{f}_{i}} \\
\mathbf{C}_{i-1} &= \mathbf{C}_{i} + \mathbf{D}_{i-1} \\
\frac{\mathrm{d} l}{\mathrm{d} \mathbf{f}_{i-1}} &= \mathbf{C}_{i-1}^\top \frac{\mathrm{d} \mathbf{a}_{i-1}}{\mathrm{d} \mathbf{x}_{i-1}} = \frac{\mathrm{d}}{\mathrm{d} \mathbf{x}_{i-1}} \bigg ( \mathbf{C}_{i-1} \cdot \mathbf{a}_{i-1} \bigg ) \\
\bigg ( \frac{\mathrm{d} l}{\mathrm{d} \sigma_{j}} \bigg )_{i-1} &= \frac{\Delta t^2}{2} \mathbf{A}_{i}^\top \frac{\mathrm{d} \mathbf{a}_{i-1}}{\mathrm{d} \sigma_{j}} = \frac{\Delta t^2}{2} \frac{\mathrm{d}}{\mathrm{d} \sigma_{j}} \bigg ( \mathbf{A}_{i} \cdot \mathbf{a}_{i-1} \bigg ) \\
\bigg ( \frac{\mathrm{d} l}{\mathrm{d} \sigma_{j}} \bigg )_{n \rightarrow i-1} &= \bigg ( \frac{\mathrm{d} l}{\mathrm{d} \sigma_{j}} \bigg )_{n \rightarrow i} + \bigg ( \frac{\mathrm{d} l}{\mathrm{d} \sigma_{j}} \bigg )_{i-1} \\
\mathbf{B}_{i-1} &= e^{-\gamma \Delta t} \mathbf{B}_{i} + \Delta t^{2} \frac{\mathrm{d} l}{\mathrm{d} \mathbf{f}_{i-1}} \\
\mathbf{A}_{i-1} &= \mathbf{A}_{i} + \mathbf{B}_{i-1} \tag{5}
\end{align*}
where $\big ( \frac{\mathrm{d} l}{\mathrm{d} \sigma_{j}} \big )_{i}$ is the contribution to $\frac{\mathrm{d} l}{\mathrm{d} \sigma_{j}}$ from step $i$ and $\big ( \frac{\mathrm{d} l}{\mathrm{d} \sigma_{j}} \big )_{n \rightarrow i}$ is the contribution to $\frac{\mathrm{d} l}{\mathrm{d} \sigma_{j}}$ from all steps from $n$ to $i$.
There are two gradient calls, in lines 3 and 4 of Equation~5.
These are vector-Jacobian products, as expected for an equivalent scheme to RAD, and consequently are efficient to compute via AD \cite{Baydin2018}.
For the simple functional forms of molecular mechanics potentials they can be coded explicitly, and hence AD is not required at all.
This is shown for the Lennard-Jones potential in the Supplementary Methods.
For MLIPs that compute potential energy and use AD to calculate the forces, second order AD can usually be used to calculate the two required gradients.

Whilst this form of the algorithm is specific to the Langevin middle integrator, the leapfrog Verlet integrator corresponds to the special case where $\gamma = 0$ ps\textsuperscript{-1}.
In this case the leading bracketed term in $\frac{\mathrm{d} l}{\mathrm{d} \mathbf{f}_{i}}$ increases to 2, 4, 6, 8, etc.\ as further steps are taken back in time (Equation~4).
This demonstrates what is known practically \cite{Ingraham2019, Metz2021, Huckelheim2023}, that gradients can explode even for a stable forward simulation.
For typical values of $\gamma = 1$ ps\textsuperscript{-1} and $\Delta t = 1$ fs the leading terms increase to 1.999, 3.996, 5.991, 7.984, etc., so gradient explosion is still a problem.
This motivates the use of gradient truncation \cite{Williams1990, Han2025}, where $\frac{\mathrm{d} l}{\mathrm{d} \mathbf{f}_{i}}$ is not accumulated beyond a certain number of reverse steps.
Here truncation was found to give more accurate gradients than gradient norm clipping \cite{Pascanu2013, Greener2024}.
The effect of gradient truncation on the accuracy of gradients is shown in Figure~\ref{fig:truncation}.
Truncation after 200 steps was used throughout the results as it appears to balance preventing gradient explosion with using information from as many steps as possible.
As described below, truncation also increases the speed of reversible simulation since reversible steps only need to be carried out whilst gradients are being accumulated.
Steps can be skipped by loading from the stored coordinates and velocities.

So far we have considered that the loss depends only on the coordinates and velocities at one point in time.
One advantage of reversible simulation over ensemble reweighting is that the loss value can take in multiple time points, for example to calculate diffusion coefficients.
In this case, additional terms are added to Equation~3 and calculated with a different set of accumulation values.
Truncation is applied separately for each.
The ability to control the gradients explicitly at every step is useful for allowing gradient truncation for losses that consider multiple time points, which would be challenging with AD software.

By carrying out the gradient calculation this way we have alleviated the problems with using RAD for DMS.
The memory cost is reduced, and hence no gradient checkpointing is required \cite{Chen2016}, since no intermediate values apart from the vectors in Equation~5 and occasional coordinate and velocity copies need to be stored.
The typical 5-10x compute overhead of RAD is reduced since we code everything explicitly.
The calculation of $\frac{\partial F(\mathbf{x}_{i}, \sigma_{j})}{\partial \sigma_{j}}$ and $\frac{\mathrm{d} F(\mathbf{x}_{i}, \sigma_{j})}{\mathrm{d} \mathbf{x}_{i}}$ each step typically takes a similar amount of time to the calculation of $F(\mathbf{x}_{i}, \sigma_{j})$, suggesting a slowdown of around 3x over the forward simulation, though for molecular mechanics force fields it is often possible to share calculations when computing the three values explicitly as shown in the Supplementary Methods.
In the absence of gradient truncation, the cost is one forward simulation followed by the reverse simulation consisting of one standard and two RAD calls to the force function.
However truncating every 200 steps, in addition to preventing gradient explosion, means that the reverse simulation only needs to take a fraction of the steps of the forward simulation depending on how often snapshots contribute to the loss.
When training the water model snapshots are taken every 2000 steps, so reversible simulation only needs to be done for a tenth of steps.
Consequently, the computation count is similar to the forward simulation and ensemble reweighting.
Concretely, on 32 CPU cores (Intel Xeon Gold 6258R) the water model with 2685 atoms runs at 2.3 ms per forward step, 3.9 ms per reverse step, and 2.7 ms per step for a 50 ps training run.
OpenMM \cite{Eastman2017} on the same machine runs at 1.2 ms per step for a standard simulation with the same parameters.

\revision{A similar derivation should yield related but different equations for other integrators and thermostats such as the Nosé-Hoover thermostat.}
Other Langevin solvers such as BAOAB splitting \cite{Leimkuhler2013} may also be suitable due to different convergence properties.
\revision{Constant pressure simulation presents more of a challenge.
The virial depends on the force field parameters, so barostats that use the virial will contribute additional terms to the gradient that are not explored here.
Monte Carlo barostats are likely incompatible with accurate gradients, as the probability of accepting the volume move depends on the force field parameters but this does not propagate through the stochastic sampling.
In the water model training we used a barostat during equilibration but not during the production run for this reason.}
Here we avoid the complexities of constrained bonds and angles, virtual sites and Ewald summation for long-range electrostatics, though the approach should extend to include them.

\subsubsection*{Implementation}

We implemented reversible simulation in the Julia language \cite{Bezanson2017} due to its flexibility, speed and growing use in science \cite{Roesch2023}.
The Molly.jl MD package \cite{Greener2024} was used for standard MD components such as neighbour lists and periodic boundary conditions.
LoopVectorization.jl and Polyester.jl were used to improve performance.
Double floating point precision was used throughout to increase numerical precision (see Figure~\ref{fig:gradients}B).
Integer random seeds were stored from the forward simulation and used to generate the same random velocities $\mathbf{n}_{i}$ during the reverse simulation.
Gradients were computed using Zygote.jl \cite{Innes2018} and Enzyme.jl \cite{Moses2020, Moses2021}.
MDAnalysis \cite{Gowers2016} and BioStructures \cite{Greener2020} were used for analysis.
Ensemble reweighting was implemented following ForceBalance \cite{Wang2014} with AD used to calculate the required $\frac{\partial l}{\partial \sigma_{j}}$ and $\frac{\mathrm{d} E}{\mathrm{d} \sigma_{j}}$ gradients for improved speed and accuracy.
The same number of snapshots were used to calculate the loss for reversible simulation and ensemble reweighting.
For the molecular mechanics models, the required force gradients were explicitly derived and implemented for bonded and non-bonded terms for all functional forms.
\revision{For MLIPs, AD was used to calculate the force gradients.}

\subsubsection*{Learning water models}

To train the water models we used a cubic box with 3 nm sides containing 895 water molecules.
The Langevin middle integrator with $\gamma = 1$ ps\textsuperscript{-1}, a temperature of 295.15 K, a time step of 1 fs, no bond or angle constraints, a 1 nm cutoff for non-bonded interactions and the reaction field approximation for long range electrostatics were used.
Each epoch an equilibrium simulation of 10 ps was followed by a production simulation of 50 ps, with the loss computed from snapshots taken every 2 ps.
A Monte Carlo barostat was used to set the pressure to 1 bar during equilibration but not during the production run.

The enthalpy of vapourisation was calculated following the procedure in OpenFF Evaluator \cite{Boothroyd2022}.
The gas potential energy was pre-computed once before training.
Since bond and angle constraints were not used during training but were used for validation simulations, 2.8 kJ/mol was added to the liquid potential energy during training as tests in OpenMM with TIP3P water indicated that not using constraints leads to this difference.
A mean squared error (MSE) loss with an experimental value of 44.12 kJ/mol was used.
The RDF was calculated for O-O and O-H distances using the differentiable procedure from Wang et al.\ 2023 \cite{Wang2023} and experimental data from Soper 2013 \cite{Soper2013}.
\revision{The RDF loss was the sum of the absolute differences between the simulated and experimental values.
The enthalpy of vapourisation and RDF losses were weighted equally.}

In addition to the Lennard-Jones or alternative parameters described below, the TIP3P starting parameters \cite{Jorgensen1983} of O partial charge -0.834, O-H bond distance 0.09572 nm, O-H bond force constant 462750 kJ mol\textsuperscript{-1} nm\textsuperscript{-2}, H-O-H angle 1.824 radians and H-O-H angle force constant 836.8 kJ/mol were used.
The Adam optimiser with a learning rate of $2 \times 10^{-3}$ was used, parameter values were divided by their starting values for optimisation to account for their different sizes, and a maximum gradient magnitude of 1000 per parameter was applied.
Training was carried out on 32 CPU cores for a week or around 1000 epochs.

Validation simulations were carried out using OpenMM \cite{Eastman2017}.
At each temperature from 260 K to 365 K at 5 K intervals, a 120 ns simulation was run with the first 20 ns being discarded as equilibration.
The Langevin middle integrator with $\gamma = 1$ ps\textsuperscript{-1}, the Monte Carlo barostat with a pressure of 1 bar, a time step of 2 fs, constrained bonds and angles, a 1 nm cutoff for non-bonded interactions and particle mesh Ewald for long range electrostatics were used.
Snapshots were saved for analysis every 50 ps.
For the self-diffusion coefficient, 5 short 5 ns equilibration simulations were run as above followed by 5 100 ps simulations in the NVE ensemble using the Verlet integrator with a time step of 1 fs.
The diffusion coefficient was calculated as described in the later section on gas diffusion.
The dielectric constant was calculated following the procedure in OpenFF Evaluator \cite{Boothroyd2022}.
The RDF was calculated using MDAnalysis \cite{Gowers2016} \revision{from a simulation at 300 K}.

\subsubsection*{Alternative functional forms}

Here we outline the potential energy functions used for the alternative functional forms.
These were only applied to the oxygen atoms in each molecule by setting $\varepsilon = 0$ kJ/mol or similar for hydrogen.
The starting O partial charge and bonded parameters are always those from TIP3P.
In each case $r$ is the interatomic distance.
The Lennard-Jones potential is standard and has parameters $\sigma$ and $\varepsilon$:
$$
V(r, \sigma, \varepsilon) = 4 \varepsilon \Bigg [ \bigg ( \frac{\sigma}{r} \bigg )^{12} - \bigg ( \frac{\sigma}{r} \bigg )^{6} \Bigg ]
$$
The TIP3P starting parameters $\sigma = 0.315$ nm and $\varepsilon = 0.636$ kJ/mol were used, and also where relevant for other functional forms.

The double exponential potential has parameters $\sigma$, $\varepsilon$, $\alpha$ and $\beta$:
$$
V(r, \sigma, \varepsilon, \alpha, \beta) = \varepsilon \left[ \frac{\beta e^\alpha}{\alpha - \beta} \exp \left( -\alpha \frac{r}{r_m} \right) - \frac{\alpha e^\beta}{\alpha - \beta} \exp \left( -\beta \frac{r}{r_m} \right) \right]
$$
where $r_m = 2^\frac{1}{6} \sigma$.
The starting values $\alpha = 16.766$ and $\beta = 4.427$ from Horton et al.\ 2023 \cite{Horton2023} were used.

The Buckingham potential has parameters $A$, $B$ and $C$:
$$
V(r, A, B, C) = A \exp(-B r) - \frac{C}{r^6}
$$
The starting values \revision{$A = 359999$ kJ/mol, $B = 37.795$ nm\textsuperscript{-1} and $C = 0.002343$ kJ/mol nm\textsuperscript{6}} were used after a fit of the three parameters to the TIP3P Lennard-Jones potential curve.

The Lennard-Jones soft core potential has parameters $\sigma$, $\varepsilon$, $\alpha$ and $\lambda$:
\begin{align*}
V(r, \sigma, \varepsilon, \alpha, \lambda) &= 4 \varepsilon \Bigg [ \bigg ( \frac{\sigma}{r_{sc}} \bigg )^{12} - \bigg ( \frac{\sigma}{r_{sc}} \bigg )^{6} \Bigg ] \\
r_{sc} &= (r^6 + \alpha \sigma^6 \lambda^p)^{1/6}
\end{align*}
where $p = 2$.
The starting values $\alpha = 0.1$ and $\lambda = 0.1$ were used.

\revision{For MACE-OFF23 the small model with Float32 precision was used \cite{Kovacs2023}.
PythonCall.jl was used to call the PyTorch MACE.
The same simulation parameters as previously except a time step of 0.5 fs and a learning rate of $1 \times 10^{-4}$ were used.
Due to the slower run time, each epoch an equilibrium simulation of 5 ps was followed by a production simulation of 10 ps.
For validation a 1.2 ns simulation at 300 K was carried out with the first 200 ps being discarded as equilibration.
Training and validation were carried out on one A100 GPU.}

\subsubsection*{Gas diffusion in water}

Unless otherwise stated, the same simulation and training options as the water model were used.
Only the non-bonded parameters were trained.
No barostat was used during equilibration.
The same box of 895 water molecules was used except 10 water molecules were randomly replaced each epoch with oxygen molecules followed by an energy minimisation.
Snapshots were taken every \revision{500} fs.
The MSD of oxygen gas molecules was calculated, accounting for the periodic boundary conditions, across multiple time segments spanning half the simulation time. This was divided by 6 times the segment time to obtain $D$ from Einstein's relation.

Training simulations were carried out using the Langevin middle integrator with $\gamma = 1$ ps\textsuperscript{-1} and a time step of 1 fs.
Training in the NVT ensemble was found to give better results than the NVE ensemble and represents a likely use case.
Consequently, the validation simulation were also run in the NVT ensemble.
The loss was the MSE to an experimental $D$ value of $2.0 \times 10^{-9}$ m\textsuperscript{2} s\textsuperscript{-1} \cite{Han1996} multiplied by $10^{18}$\revision{, plus the enthalpy of vapourisation and RDF losses from the water model multiplied by 0.05.}
Starting parameters for the oxygen gas of $\sigma = 0.3297$ nm and $\varepsilon = 0.438$ kJ/mol were taken from Wang et al.\ 2021 \cite{Wang2021}.
The Adam optimiser with a learning rate of $5 \times 10^{-4}$ was used.
For validation, 5 simulations of 100 ps were run after separate 10 ps equilibration runs and the $D$ value averaged.
\revision{The bulk water validation simulations were the same as for the water model and did not contain oxygen molecules.}

\subsubsection*{Neural network model for diamond}

The Stillinger-Weber prior \cite{Stillinger1985} was implemented in Julia.
The starting parameters were those for silicon with modified length and energy scales $\sigma_{SW} = 0.14$ nm and $\varepsilon_{SW} = 200$ kJ/mol to account for the smaller carbon atom \cite{Thaler2021}.
Rather than implement the DimeNet++ model \cite{Gasteiger2020} in Julia, PythonCall.jl was used to call the Jax code from Thaler and Zavadlav 2021 \cite{Thaler2021, Schoenholz2019} on \revision{the} GPU.
In this section the notation from that paper is matched.
A cubic box with 1.784 nm sides containing 1000 carbon atoms was used, representing 5 diamond unit cells in each direction.
The Langevin middle integrator with $\gamma = 4$ ps\textsuperscript{-1}, a temperature of 298 K and a time step of 0.5 fs were used.
The loss was defined as:
$$
l = \frac{\gamma_{\sigma}}{9} \sum_{i,j} \sigma_{ij}^2 + \frac{\gamma_C}{3} \left ( (C_{11} - \tilde{C}_{11})^2 + (C_{12} - \tilde{C}_{12})^2 + (C_{44} - \tilde{C}_{44})^2 \right )
$$
where $\gamma_{\sigma} = 5 \times 10^{-8}$ kJ\textsuperscript{-2} mol\textsuperscript{2} nm\textsuperscript{6}, $\gamma_C = 10^{-10}$ kJ\textsuperscript{-2} mol\textsuperscript{2} nm\textsuperscript{6}, $\tilde{C}_{11} = 1079$ GPa, $\tilde{C}_{12} = 124$ GPa and $\tilde{C}_{44} = 578$ GPa.
The crystal is assumed to have zero stress for vanishing strain $\boldsymbol\epsilon = \boldsymbol0$.
The virial stress tensor $\boldsymbol\sigma^V$ is calculated \cite{Chen2020} as:
$$
\boldsymbol\sigma^V = \frac{1}{\Omega} \left [ - \sum_{k=1}^{N} m_k \mathbf{v}_k \otimes \mathbf{v}_k - \mathbf{F}^T \mathbf{R} + \left ( \frac{\partial U}{\partial \mathbf{h}} \right )^T \mathbf{h} \right ]
$$
where $N$ is the number of atoms, $\otimes$ is the outer product, $m_k$ are the atomic masses, $\mathbf{v}_k$ are the atom velocities, $\mathbf{R}$ is the atomic coordinate array ($N \times 3$), $\mathbf{F}$ is the atomic force array ($N \times 3$), $U$ is the potential energy, $\mathbf{h}$ is the lattice tensor describing the simulation box and $\Omega = \det(\mathbf{h})$ is the box volume.
The isothermal elastic stiffness tensor $\mathbf{C}$ was calculated at constant strain $\boldsymbol\epsilon$ via the stress fluctuation method:
$$
C_{ijkl} = \frac{\partial \left< \sigma_{ij}^V \right>}{\partial \epsilon_{kl}} = \left< C_{ijkl}^B \right> + \frac{N k_B T}{\Omega} \left ( \delta_{ik} \delta_{jl} + \delta_{il} \delta_{jk} \right )
$$
with $C_{ijkl}^B = \frac{1}{\Omega} \frac{\partial^2 U}{\partial \epsilon_{ij} \epsilon_{kl}}$ and Kronecker delta $\delta_{ij}$.
Second order AD was used to calculate $\frac{\partial^2 U}{\partial \epsilon_{ij} \epsilon_{kl}}$, meaning that third order AD was used to calculate the gradient of the loss function.
$C_{11}$, $C_{12}$ and $C_{44}$ were calculated from $\mathbf{C}$ \cite{Thaler2021}.
The Born contribution to the stress tensor is omitted as it is difficult to calculate with reversible simulation and it is a considerably smaller term than the others.
The loss was computed from snapshots taken every 250 fs.
AD was used in Julia or Jax to compute the required derivatives.
The training simulation time was scaled up over epochs, and was set to 0.5 fs multiplied by the epoch number with no equilibration.
By the end of training at 2000 epochs the simulation time was 1 ps, which was found to be sufficient for learning.
The Adam optimiser with a learning rate of $2 \times 10^{-3}$ for the DimeNet++ parameters and $5 \times 10^{-4}$ for the Stillinger-Weber parameters was used.
The validation simulations with the learned model were 100 ps.
Training and validation were carried out on one A100 GPU.
Other details are the same as Thaler and Zavadlav 2021 \cite{Thaler2021}.

\subsection*{Availability}

\revision{Training scripts, validation scripts and trained models} are available under a permissive licence at \url{https://github.com/greener-group/rev-sim}.
Molly.jl is available at \url{https://github.com/JuliaMolSim/Molly.jl}.

\subsection*{Conflict of interest}

The author declares no competing interests.

\subsection*{Acknowledgements}

I thank the Sjors Scheres group, Stephan Thaler, Josh Fass, Yutong Zhao, Yuanqing Wang, Daniel Cole, Joshua Horton, Kresten Lindorff-Larsen, Patrick Kidger and James Foster for useful discussions; all contributors to Molly.jl; William Moses and Valentin Churavy for support with Enzyme.jl; and Jake Grimmett, Toby Darling and Ivan Clayson for help with high-performance computing.
This work was supported by the Medical Research Council, as part of United Kingdom Research and Innovation (also known as UK Research and Innovation) [MC\_UP\_1201/33].
For the purpose of open access, the MRC Laboratory of Molecular Biology has applied a CC BY public copyright licence to any Author Accepted Manuscript version arising.

\begin{footnotesize}
\bibliographystyle{unsrtnat}
\bibliography{library}
\end{footnotesize}

\clearpage
\setcounter{page}{1}

\begin{center}
\section*{Reversible molecular simulation for training classical and machine learning force fields}
\subsection*{Joe G Greener}
\section*{Supplementary Methods and Data}
\end{center}

\setcounter{figure}{0}
\setcounter{table}{0}
\renewcommand{\thefigure}{S\arabic{figure}}
\renewcommand{\thetable}{S\arabic{table}}

\vspace{10px}

\begin{figure}[h]
  \centering
  \includegraphics[width=1.0\textwidth]{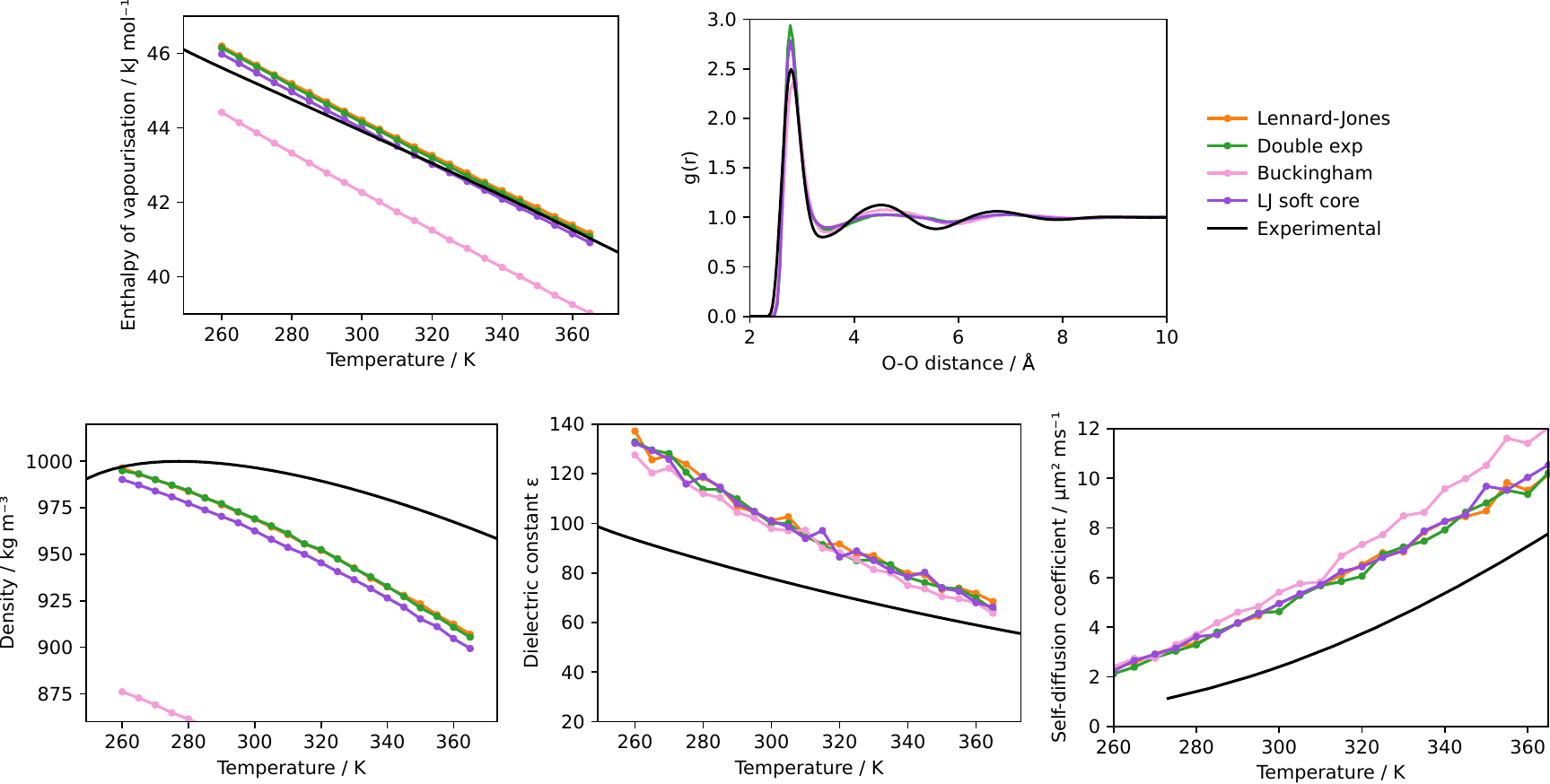}
  \caption{\revision{Validating oxygen diffusion models on condensed phase properties of bulk water with no oxygen molecules present. See Figure~\ref{fig:water_val} for more details.}}
  \label{fig:diff_val}
\end{figure}

\begin{figure}[h]
  \centering
  \includegraphics[width=1.0\textwidth]{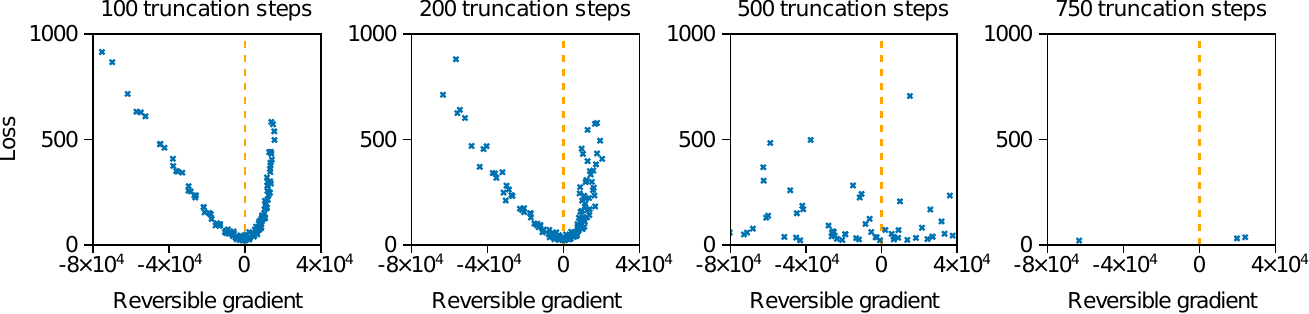}
  \caption{The effect of the number of steps after which gradients are truncated on gradient accuracy. A simulation of 1000 steps with a single loss snapshot was run with various parameters as described in Figure~\ref{fig:gradients}. The gradients were truncated after a given number of steps and the losses are shown plotted against the gradients. Some values are outside the range shown. 200 truncation steps were used throughout the results.}
  \label{fig:truncation}
\end{figure}

\subsection*{Reversible molecular simulation derivation}

Consider a single simulation step of the Langevin middle integrator as shown in Equation~2.
One operation at a time can be explicitly computed with constants represented by values of $k$ and intermediate vectors represented by values of $\mathbf{w}$:
\begin{align*}
k_{1} &= \frac{\Delta t}{2} \\
k_{2} &= e^{-\gamma \Delta t} \\
k_{3} &= \sqrt{1 - k_{2}^{2}} \\
\mathbf{f}_{1} &= F(\mathbf{x}_{1}, \sigma_{j}) \\
\mathbf{a}_{1}^{'} &= \mathbf{f}_{1} / \textbf{m} \\
\mathbf{w}_{1} &= \mathbf{a}_{1}^{'} \Delta t \\
\mathbf{w}_{2} &= \mathbf{v}_{0+1/2} + \mathbf{w}_{1} \\
\mathbf{w}_{3} &= k_{1} \mathbf{w}_{2} \\
\mathbf{w}_{4} &= \mathbf{x}_{1} + \mathbf{w}_{3} \\
\mathbf{w}_{5} &= k_{3} \mathbf{n}_{1} \\
\mathbf{w}_{6} &= k_{2} \mathbf{w}_{2} \\
\mathbf{v}_{1+1/2} &= \mathbf{w}_{5} + \mathbf{w}_{6} \\
\mathbf{w}_{7} &= k_{1} \mathbf{v}_{1+1/2} \\
\mathbf{x}_{2} &= \mathbf{w}_{4} + \mathbf{w}_{7} \\
l &= L(\mathbf{x}_{2}, \mathbf{v}_{1+1/2}, \sigma_{j})
\end{align*}
The multi-variable chain rule can then be used to compute $\frac{\mathrm{d} l}{\mathrm{d} \sigma_{j}}$:
\begin{align*}
\frac{\mathrm{d} l}{\mathrm{d} \sigma_{j}} &= \frac{\partial l}{\partial \sigma_{j}} + \frac{\mathrm{d} l}{\mathrm{d} \mathbf{f}_{1}}^\top \frac{\partial F(\mathbf{x}_{1}, \sigma_{j})}{\partial \sigma_{j}} \\
\frac{\mathrm{d} l}{\mathrm{d} \mathbf{f}_{1}} &= \frac{\mathrm{d} l}{\mathrm{d} \mathbf{a}_{1}^{'}}^\top \frac{\mathrm{d} \mathbf{a}_{1}^{'}}{\mathrm{d} \mathbf{f}_{1}} = \frac{1}{\mathbf{m}} \frac{\mathrm{d} l}{\mathrm{d} \mathbf{w}_{1}}^\top \frac{\mathrm{d} \mathbf{w}_{1}}{\mathrm{d} \mathbf{a}_{1}^{'}} = \frac{\Delta t}{\mathbf{m}} \frac{\mathrm{d} l}{\mathrm{d} \mathbf{w}_{2}}^\top \frac{\mathrm{d} \mathbf{w}_{2}}{\mathrm{d} \mathbf{w}_{1}} = \frac{\Delta t}{\mathbf{m}} \bigg ( \frac{\mathrm{d} l}{\mathrm{d} \mathbf{w}_{3}}^\top \frac{\mathrm{d} \mathbf{w}_{3}}{\mathrm{d} \mathbf{w}_{2}} + \frac{\mathrm{d} l}{\mathrm{d} \mathbf{w}_{6}}^\top \frac{\mathrm{d} \mathbf{w}_{6}}{\mathrm{d} \mathbf{w}_{2}} \bigg ) \\
&= \frac{\Delta t}{\mathbf{m}} \bigg ( \frac{\Delta t}{2} \frac{\mathrm{d} l}{\mathrm{d} \mathbf{w}_{4}} + e^{-\gamma \Delta t} \frac{\mathrm{d} l}{\mathrm{d} \mathbf{v}_{1+1/2}} \bigg ) = \frac{\Delta t}{\mathbf{m}} \bigg ( \frac{\Delta t}{2} \frac{\mathrm{d} l}{\mathrm{d} \mathbf{x}_{2}} + e^{-\gamma \Delta t} \frac{\mathrm{d} l}{\mathrm{d} \mathbf{w}_{7}}^\top \frac{\mathrm{d} \mathbf{w}_{7}}{\mathrm{d} \mathbf{v}_{1+1/2}} + e^{-\gamma \Delta t} \frac{\mathrm{d} l}{\mathrm{d} \mathbf{v}_{1+1/2}} \bigg ) \\
&= \frac{\Delta t^{2}}{2 \mathbf{m}} \bigg ( 1 + e^{-\gamma \Delta t} \bigg ) \frac{\mathrm{d} l}{\mathrm{d} \mathbf{x}_{2}} + \frac{\Delta t}{\mathbf{m}} e^{-\gamma \Delta t} \frac{\mathrm{d} l}{\mathrm{d} \mathbf{v}_{1+1/2}}
\end{align*}
This is the first term in Equation~4.
Following a similar process with assistance from Symbolics.jl \cite{Gowda2022} further terms, which quickly increase in complexity, can be derived.
Examining the relationship between these terms manually leads to the relations in Equation~5.

\subsection*{Force gradients}

The Lennard-Jones potential between two atoms is defined by potential energy $V$ for interatomic distance $r$ and atom pair parameters $\sigma$ and $\varepsilon$.
The magnitude of the force $F$ and the gradients required for reversible simulation are given by:
\begin{align*}
V(r, \sigma, \varepsilon) &= 4 \varepsilon \Bigg [ \bigg ( \frac{\sigma}{r} \bigg )^{12} - \bigg ( \frac{\sigma}{r} \bigg )^{6} \Bigg ] \\
F(r, \sigma, \varepsilon) = - \frac{\mathrm{d} V(r, \sigma, \varepsilon)}{\mathrm{d} r} &= \frac{24 \varepsilon}{r} \Bigg [ 2 \bigg ( \frac{\sigma}{r} \bigg )^{12} - \bigg ( \frac{\sigma}{r} \bigg )^{6} \Bigg ] \\
\frac{\mathrm{d} F(r, \sigma, \varepsilon)}{\mathrm{d} r} &= -\frac{24 \varepsilon}{r^{2}} \Bigg [ 26 \bigg ( \frac{\sigma}{r} \bigg )^{12} - 7 \bigg ( \frac{\sigma}{r} \bigg )^{6} \Bigg ] \\
\frac{\mathrm{d} F(r, \sigma, \varepsilon)}{\mathrm{d} \sigma} &= \frac{144 \varepsilon}{r \sigma} \Bigg [ 4 \bigg ( \frac{\sigma}{r} \bigg )^{12} - \bigg ( \frac{\sigma}{r} \bigg )^{6} \Bigg ] \\
\frac{\mathrm{d} F(r, \sigma, \varepsilon)}{\mathrm{d} \varepsilon} &= \frac{24}{r} \Bigg [ 2 \bigg ( \frac{\sigma}{r} \bigg )^{12} - \bigg ( \frac{\sigma}{r} \bigg )^{6} \Bigg ]
\end{align*}
Significant computation can be reused when calculating these quantities.
Note that when $r \gg \sigma$ the power 12 term will approach zero and $\frac{\mathrm{d} F}{\mathrm{d} \sigma}$ and $\frac{\mathrm{d} F}{\mathrm{d} \varepsilon}$ will have the same sign.

\subsection*{Comparison to ensemble reweighting}

Consider for example the ForceBalance approach \cite{Wang2014, Wang2013, Wang2013b}.
If $l$ is a generic thermodynamic average property then:
\begin{align*}
\langle l \rangle &= \sum_{i=1}^{M} l(\mathbf{x}_{i}, \sigma_{j}) p_{i}(\mathbf{x}_{i}, \sigma_{j}) \\
&= \sum_{i=1}^{M} \frac{l(\mathbf{x}_{i}, \sigma_{j}) \exp \Big ( -\frac{E_{i}(\mathbf{x}_{i}, \sigma_{j})}{k_{B} T} \Big ) }{Q} \\
Q &= \sum_{k=1}^{M} \exp \bigg ( -\frac{E_{k}(\mathbf{x}_{k}, \sigma_{j})}{k_{B} T} \bigg )
\end{align*}
where $M$ is the number of microstates, $p_{i}$ is the probability of state $i$, $E_{i}$ is the potential energy of state $i$, $Q$ is the partition function and the angle brackets represent the average over microstates.
By differentiating this \cite{Wang2014, Wang2013, Wang2013b} we obtain:
\begin{align*}
\frac{\mathrm{d} \langle l \rangle}{\mathrm{d} \sigma_{j}} &= \left\langle \frac{\partial l}{\partial \sigma_{j}} \right\rangle + \frac{1}{k_{B} T} \bigg ( \left\langle l \frac{\mathrm{d} E}{\mathrm{d} \sigma_{j}} \right\rangle - \langle l \rangle \left\langle \frac{\mathrm{d} E}{\mathrm{d} \sigma_{j}} \right\rangle \bigg )
\end{align*}
This can be compared to Equation~1.
Finite differences can be used to calculate $\frac{\partial l}{\partial \sigma_{j}}$ and $\frac{\mathrm{d} E}{\mathrm{d} \sigma_{j}}$ \cite{Wang2014}, but AD provides a way to do this faster and with higher accuracy \cite{Thaler2021}.
Typically, one or more simulations are run and the snapshots sampled are taken as representative of the microstates.
This assumes sufficient sampling of low energy regions and requires enough time between snapshots to reduce correlation.
The first term is the same as in Equation~1 and represents the direct dependence of $l$ on the parameters.
The second term represents how a change in the parameters affects the weighting of states in the ensemble.
Reversible simulation does this by differentiating through a simulation, whereas the ensemble reweighting approach reweights the snapshots based on how the potential energy depends on the parameters.
Ensemble reweighting therefore only consider snapshot states, whereas reversible simulation considers a number of steps prior to each snapshot state depending on gradient truncation.
Since reordering states does not change the gradients arising from ensemble reweighting, observables that depend on multiple time points such as diffusion coefficients are not directly applicable to this scheme.
DiffTRe extends the above approach by using thermodynamic perturbation theory to reuse states, allowing for more efficient training \cite{Thaler2021}.

\subsection*{Comparison to the adjoint method}

The adjoint method differentiates an ordinary differential equation (ODE) before discretising it \cite{Chen2018, Kidger2021}.
Consider a loss function $L$ whose input is the result of an ODE solver acting on hidden state $\mathbf{z}$:
$$
l = L(\mathbf{z}(t_{1})) = L \bigg ( \mathbf{z}(t_{0}) + \int_{t_{0}}^{t_{1}} f(\mathbf{z}(t), \sigma_{j}) dt \bigg )
$$
The adjoint $\mathbf{a}(t)$ determines the gradient of the loss with respect to $\mathbf{z}(t)$:
$$
\mathbf{a}(t) = \frac{\partial l}{\partial \mathbf{z}(t)}
$$
It can then be shown \cite{Chen2018} that:
\begin{align*}
\frac{\mathrm{d} \mathbf{a}(t)}{\mathrm{d} t} &= -\mathbf{a}(t)^\top \frac{\partial f(\mathbf{z}(t), \sigma_{j})}{\partial \mathbf{z}(t)} \\
\frac{\mathrm{d} l}{\mathrm{d} \sigma_{j}} &= -\int_{t_{1}}^{t_{0}} \mathbf{a}(t)^\top \frac{\partial f(\mathbf{z}(t), \sigma_{j})}{\partial \sigma_{j}} dt
\end{align*}
The required integrals for solving $\mathbf{z}$, $\mathbf{a}$ and $\frac{\mathrm{d} l}{\mathrm{d} \sigma_{j}}$ can be computed in a single call to an ODE solver.
This steps back through time starting from the final state, similar to reversible simulation.
The two vector-Jacobian products above are similar to the two in Equation~5.
However, reversible simulation discretises the differential equation before differentiating it \cite{Kidger2021}.
This means that the gradients match those of the forward simulation to within numerical error.
By contrast, the adjoint method solves a different equation to obtain the gradients, which can cause problems \cite{Gholami2019, Onken2020}.
It can be unclear how to best solve this adjoint equation.
The forward simulation is stable for conventional MD cases, but this is not guaranteed for the adjoint equation \cite{Kim2021}, so it makes sense to use the gradients of the forward simulation if possible.
There has also been work on second order neural ODEs \cite{Norcliffe2020}.


\end{document}